\documentclass[lettersize,journal]{IEEEtran}
\usepackage{amsmath,amsfonts}
\usepackage{enumerate}
\usepackage{algorithmic}
\usepackage{algorithm}
\usepackage{array}
\usepackage[caption=false,font=normalsize,labelfont=sf,textfont=sf]{subfig}
\usepackage{textcomp}
\usepackage{stfloats}
\usepackage{url}
\usepackage{verbatim}
\usepackage{graphicx}
\usepackage{cite}
\usepackage{booktabs} 
\usepackage{float}  
\usepackage{subfloat}  
\begin{document}
	
	\title{Speech Emotion Recognition Via CNN-Transformer and Multidimensional Attention Mechanism}
	
\author{
	Xiaoyu Tang,
	Yixin Lin,
	Ting Dang,
	Yuanfang Zhang,
	Jintao Cheng
	\thanks{Corresponding author: Xiaoyu Tang. E-mail address: tangxy@scnu.edu.cn.} 
	\thanks{Xiaoyu Tang is with the School of Electronic and Information Engineering, Faculty of Engineering, South China Normal University, Foshan, Guangdong 528225, China, and also with the School of Physics and Telecommunications Engineering, South China Normal University, Guangzhou, Guangdong 510000, China.
		
		Yixin Lin is with the School of Electronic and Information Engineering, Faculty of Engineering, South China Normal University, Foshan, Guangdong 528225, China.
		
		Ting Dang is with the Nokia Bell Labs, Cambridge, UK.
		
		Yuanfang Zhang is with the Autocity (Shenzhen) Autonomous Driving Co.,ltd.
		
		Jintao Cheng is with the School of Physics and Telecommunications Engineering, South China Normal University, Guangzhou, Guangdong 510000, China.
}}
	
	\markboth{Journal of \LaTeX\ Class Files,~Vol.~14, No.~8, August~2021}%
	{Shell \MakeLowercase{\textit{et al.}}: A Sample Article Using IEEEtran.cls for IEEE Journals}
	
	\IEEEpubid{0000--0000/00\$00.00~\copyright~2021 IEEE}
	
	\maketitle
	
	\begin{abstract}
		Speech Emotion Recognition (SER) is crucial in human-machine interactions. Mainstream approaches utilize Convolutional Neural Networks or Recurrent Neural Networks to learn local energy feature representations of speech segments from speech information, but struggle with capturing global information such as the duration of energy in speech. Some use Transformers to capture global information, but there is room for improvement in terms of parameter count and performance. Furthermore, existing attention mechanisms focus on spatial or channel dimensions, hindering learning of important temporal information in speech. In this paper, to model local and global information at different levels of granularity in speech and capture temporal, spatial and channel dependencies in speech signals, we propose a Speech Emotion Recognition network based on CNN-Transformer and multi-dimensional attention mechanisms. Specifically, a stack of CNN blocks is dedicated to capturing local information in speech from a time-frequency perspective. In addition, a time-channel-space attention mechanism is used to enhance features across three dimensions. Moreover, we model local and global dependencies of feature sequences using large convolutional kernels with depthwise separable convolutions and lightweight Transformer modules. We evaluate the proposed method on IEMOCAP and Emo-DB datasets and show our approach significantly improves the performance over the state-of-the-art methods\footnote{Our code is available on https://github.com/SCNU-RISLAB/CNN-Transformer-and-Multidimensional-Attention-Mechanism}.
	\end{abstract}
	
	
	\section{Introduction}
	\IEEEPARstart{E}{motion} recognition has significant importance in various fields, especially in increasingly common human-computer interaction systems \cite{el2011survey}, and speech emotion recognition (SER) has promising applications in areas such as mental health monitoring \cite{low2010detection}, educational assistance, personalized content recommendation, and customer service quality monitoring. Speech contains rich emotional information, and as one of the most basic human communication methods, speech emotion recognition is particularly important for computers to analyze and respond to the emotional state of human users and respond to them accordingly. With the rapid development of artificial intelligence, speech emotion recognition has received extensive research attention. Human speech contains a wealth of information, including not only the language content but also attributes such as gender and emotions of the speaker. It is of great significance to accurately identify emotional information from speech signals.
	
	Feature extraction of speech is a rather important and challenging task in speech emotion recognition, and the extraction of features directly affects the effectiveness of subsequent model training and the accuracy of the final algorithm for emotion recognition. Speech features can be categorized as acoustic-based features and deep learning-based features where acoustic-based features can be broadly classified into rhythmic features \cite{mahdhaoui2008motherese}, spectral-based correlation features \cite{bou2000comparative}, and phonetic features \cite{gobl2003role}. Among them, spectral-based correlation features reflect the characteristics of the signal in the frequency domain, where there are differences in the performance of different emotions in the frequency domain. Based on the spectral correlation features include linear spectrum \cite{hernando1997linear} and inverse spectrum \cite{barpanda2019iris}, Linear Prediction Cofficients (LPC), Log Frequency Power Coefficients (LFPC), etc.; Inverse spectrum includes Mel-Frequency Cepstrum Coefficients (MFCC), Linear Prediction Cepstrum Cofficients (LPCC), etc. Among them, MFCC is regarded as a low-level feature based on human knowledge, which is widely used in the field of speech.
	
	\IEEEpubidadjcol
	
	Early SER algorithms mainly used acoustic-based features and combined with traditional machine learning algorithms, including hidden Markov models \cite{qin2011hmm}, Gaussian mixture models \cite{pribil2019artefact}, and support vector machines \cite{schuller2010cross}. In recent years, deep learning-based neural networks have gradually become active in the field of speech emotion recognition \cite{gupta2019deep ,kumar2019comparison}, and compared with traditional models, deep learning-based models have shown better performance in speech emotion recognition. Deep learning-based features use neural networks to learn more advanced features from the original signal of speech or some low-dimensional acoustic features. Convolutional neural networks (CNNs) are effective in capturing local acoustic details in speech, while long short-term memory networks (LSTMs) are widely used in speech emotion recognition for modeling dynamic information and temporal dependencies in speech. Additionally, attention mechanisms are also a key factor in improving model performance, as they can adaptively focus on the importance of different features to obtain better speech features at the discourse level. For example, Qi \textit{et al.} \cite{cao2021hierarchical} proposed a hierarchical network based on static and dynamic features, which uses LSTM to encode dynamic and static features of speech and designs a gating model to fuse the features, an attention mechanism is used to acquire the discourse-level speech features. Liu \textit{et al.} \cite{liu2020speech} proposed a local-global perceptual depth representation learning system. One module contains a multiscale CNN and a time-frequency CNN (TFCNN) to learn the local representation, and in the other module, a Capsule network with an improved routing algorithm is utilized to design a multi-block dense connection structure, which can learn both shallow and deep global information.
	
	Although speech emotion recognition (SER) models composed of CNNs exhibit better performance than traditional models, these networks can only extract local information in speech, such as the energy and rhythm of a particular segment of speech, while struggling to learn global information in speech features, such as the overall volume and speaking rate of the speaker, and the duration of energy, thus neglecting the global correlation of features\cite{khalil2019speech}. In recent years, the transformer \cite{vaswani2017attention} based on the self-attentive mechanism has been widely used in major deep learning tasks. Tarantino \textit{et al.} \cite{tarantino2019self} used transformer in combination with global windowing for speech emotion recognition and achieved better performance, but transformer is weak for local feature extraction. Some recent work has attempted to combine CNN and transformer to alleviate the limitations of using CNN and transformer alone. Wang \textit{et al.} \cite{wang2021novel} stacked the transformer blocks after the CNN model to improve the global features of aggregation. A model combining the transformer and CNN is proposed in \cite{hu2022multiple}, enabling it to learn local information while capturing global dependencies. The original transformer tends to have a high number of parameters for computing multi-headed self-attention, which requires a lot of resources and poses some difficulties in the training of the network. In addition, this kind of work tends to stack transformers in the last part of the model or a simple combination of transformer and CNN, which makes it hard to obtain better local information of the speech.
	
	In addition, attention mechanisms improve the effectiveness of task processing by selectively attending to features that are most relevant to the current task, and have received widespread attention in major fields. In recent years, several researchers have utilized deep learning methods for feature extraction and used attention mechanisms to improve performance. A lightweight self-attention module is proposed in \cite{kwon2021att}, which uses MLP to extract channel information and a large perceptual field extended CNN to extract spatial contextual information. Guo \textit{et al.} \cite{guo2021representation} proposed an attention mechanism based on time, frequency, and CNN channels to improve representation learning ability. However, temporal information is often embedded in speech, which reflects the dynamic changes of speech, such as pitch and energy variations over time. Temporal features can reflect the temporal context and evolution of emotion expression in speech. However, it falls short in capturing the temporal information present in speech, which represents the dynamics of speech and plays a crucial role in emotion recognition. This limitation is a common issue in both MLPs and CNNs.
	
	In this paper, we investigate how to effectively combine transformer and CNN and apply them to SER to characterize local features and global features in speech signals, and propose the temporal-channel-space attention mechanism in the model for multiple dimensions of feature enhancement. Specifically, we first use a set of stacked CNNs to capture local information in speech and learn shallow features of speech for the transformer module for better training of the transformer module. In the stacked CNN module, two sets of convolutional filters of different shapes are used to capture both temporal and frequency domain contextual information. Specifically, after stacking the CNN modules, we introduced a temporal-channel-space attention mechanism that models the contextual emotional expression of features over time, and efficiently fuses the attention of the spatial and channel dimensions of the speech feature map through the Shuffle unit. Furthermore, a combination of transformer and CNN is used to model the local and global dependencies of feature sequences by a deep separable convolution with residuals and a lightweight transformer module. The main contributions of this work are summarized as follows:
	
	\begin{itemize}
		\item A framework based on CNN and transformer is proposed for speech emotion recognition. Our framework uses time-frequency domain convolution and stacked convolution blocks to extract initial local features of speech and stacked CNN and transformer blocks are used to enhance local and global features.  
		\item To enhance the finiteness of the feature map and model the temporal information of speech, a temporal-channel-space attention mechanism called Time-Shuffle Attention (T-Sa) is used in our model. T-Sa enhances the feature map in multi-dimensions.
		\item We propose a module based on deeply separable convolution and a lightweight transformer called Lightweight convolution transformer(LCT). This model employs lightweight convolutional blocks to efficiently extract local information from features, and incorporates Coordinate Attention (CA) into the multi-head self-attention mechanism to capture long-range dependencies among features while enhancing their temporal and spectral information.
		\item Extensive experiments of our proposed model on IEMOCAP and EMO-DB datasets demonstrate the effectiveness of the model in SER tasks.
	\end{itemize}
	
	The rest of the paper is organized as follows. Section II briefly reviews related work. Details of the system are presented in Section II. In Section IV,we present experimental results to showcase the effectiveness of our model on two widely-used benchmark datasets. Section V concludes this work.
	
	\section{RELATED WORK}
	In this section, we will briefly review the algorithms related to speech emotion recognition, namely convolutional and recurrent neural networks, attention mechanism, and transformer.
	\subsection{Convolutional neural networks and recurrent neural networks}
	Speech is a continuous time-series signal, and CNN and RNN have been two main network structures for SER. Motivated by the studies of CNN in computer vision, AlexNet \cite{krizhevsky2017imagenet} and ResNet \cite{he2016deep} show promising results in image classification tasks and therefore have been studies in SER. Zhu \textit{et al.} \cite{zhu2022speech} has proposed a new Global Aware Multi-scale (GLAM) neural network that utilizes a global perception fusion module to learn multi-scale feature representations, with a focus on emotional information.  The multi-time-scale (MTS) method was introduced in \cite{guizzo2020multi}, which extends the CNNs by scaling and resampling the convolutional kernel along the time axis to increase temporal flexibility. Liu \textit{et al.} \cite{liu2020speech} proposed a local global-aware deep representation learning system that uses CNNs and Capsule Networks to learn local and deep global information.
	
	RNN can model the temporal information in speech and capture long-term dependencies in the speech signal more effectively. A new layered network HNSD was proposed \cite{cao2021hierarchical} that can efficiently integrate static and dynamic features of SER, which uses LSTM to encode static and dynamic features and gated multi-features unit (GMU) for frame level feature fusion for the emotional intermediate representation. Xu \textit{et al.} \cite{xu2020hgfm} proposed a hierarchical grained and feature model (HGFM) that uses recurrent neural networks to process both discourse-level and frame-level information of the speech. Since convolutional neural networks can capture local information of features, while recurrent neural networks take advantage of modeling temporal information, many works have combined these two approaches and achieved outstanding results. Li \cite{li2021robotic} extracted location information from MFCC features and VGGish features by bi-direction long short time memory (BiLSTM) neural network, and then fused these two features to predict emotions. Liu \textit{et al.} \cite{liu2021speech} combined triplet loss and CNN-LSTM models to obtain more discriminative sentiment information, and the proposed framework yielded excellent results in experiments. Zou \cite{zou2022speech} \textit{et al.} used CNN, BiLSTM, and wav2vec2 to extract different levels of speech information, including MFCC, spectrogram, and acoustic information, and fused these three features by an attention mechanism. Zhang \cite{zhang2019spontaneous} \textit{et al.} used CNNs to learn segment-level features in spectrograms, using a deep LSTM model to capture temporal dependencies between speech segments.
	\subsection{Attention mechanism}
	In recent years, attention mechanisms have received a lot of attention in major fields to improve the effectiveness of task processing by focusing on information that is more critical to the current task among the many inputs. A channel attention mechanism called Squeeze-and-Excitation (SE) was proposed in \cite{hu2018senet}, which assigns weights to individual channels and adaptively recalibrates the feature responses of the channels. Woo \textit{et al.} \cite{woo2018cbam} proposed a convolutional block attention module that combines both spatial and channel dimensions to obtain attention with better results. In addition, some researchers have used deep learning methods for feature extraction of speech and enhancement of feature maps using attention mechanisms. An attention pooling-based approach was proposed in \cite{li2018attention}, compared to existing average and maximum pooling, it can combine both class-agnostic bottom-up attention maps and class-specific top-down attention maps in an effective manner. Mustaqeem \textit{et al.} \cite{kwon2021att} proposed a self-attentive module (SAM) for SER systems,which uses a multilayer perceptron (MLP) to recognize global information of the channels and identifies spatial information using a special dilated CNN to generate an attentional map for both channels. SAM significantly reduces the computational and parametric overhead. A spectro-temporal-channel (STC) attention mechanism was proposed in \cite{guo2021representation}, which acquires attention feature maps along three dimensions: time, frequency, and channel. Xi \textit{et al.} \cite{xi2022frontend} employed an attention mechanism based on the time and frequency domain to introduce long-distance contextual information.
	
	The current attention mechanisms typically focus more on spatial or channel information in feature maps, often neglecting the temporal characteristics in speech. However, temporal features in speech are equally important for emotion recognition. Therefore, it is necessary to pay more attention to temporal information in attention mechanisms to better explore and utilize the temporal characteristics in speech signals.
	
	\begin{figure*}[htbp]
		\centering
		\includegraphics[scale=0.069]{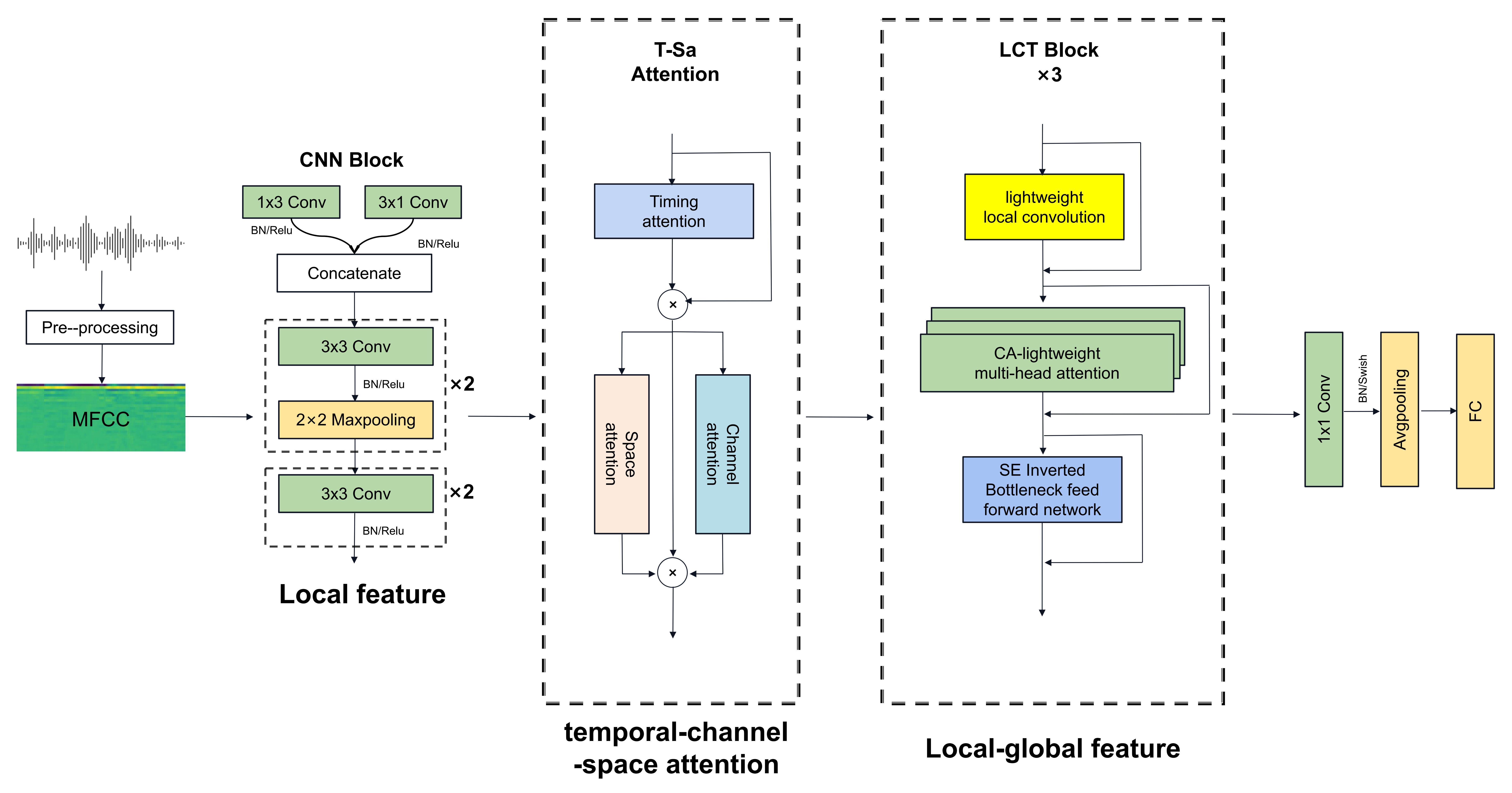}
		\caption{An illustration of our proposed speech emotion recognition framework consisting of three modules: i) CNN Block is used to extract local information in speech; ii) T-Sa attention module enhances speech information in three dimensions: time-space-channel; iii) LCT Block combines local and global information in speech.}
	\end{figure*}
	
	\subsection{Transformer}
	Transformer has been rapidly developing in the field of natural language processing (NLP) in recent years and has achieved great success. Due to its powerful ability to obtain global information, the transformer has been gradually extended to the fields of speech. An end-to-end speech emotion recognition model \cite{wang2021novel} was proposed to enhance the global feature representation of the model by using stacked transformer blocks at the end of the model. Hu \textit{et al.} \cite{hu2022multiple} took advantage of multiple models, improved the learning ability of the model by residual BLSTM, and proposed a convolutional neural network and E-transformer module to learn both local and global information. Recently, transformer-based self-supervised methods have also been applied to speech, and some transformer-based models have achieved great success in automatic speech recognition (ASR), including wav2vec \cite{schneider2019wav2vec}, VQ-wav2vec \cite{baevski2019vq}, and wav2vec2.0 \cite{baevski2020wav2vec}. There is also some work in speech emotion recognition that employs these models for migration learning. A pre-trained wav2vec2.0 model \cite{pepino2021emotion} is used as the input to the network and the outputs of multiple network layers of the pre-trained model are combined to produce a richer representation of speech features. Cai \textit{et al.} \cite{cai2021speech} proposed a multi-task learning (MTL) framework that uses the wav2vec2.0 model for feature extraction and simultaneously training for speech emotion classification and text recognition. Among computer vision tasks, ViT \cite{dosovitskiy2020image} first applied transformer directly to image patch sequences which is groundbreaking in applying transformer structure to computer vision. ViT has a superior structure and reduced computational resource consumption compared to convolutional neural networks. There are many similar approaches in the field of speech. ViT was introduced to speech and improved based on the properties of spectrograms in \cite{ristea2203septr}, which proposed a separable transformer (SepTr) that uses the transformer to process tokens at the same time and the same frequency interval, respectively. In \cite{xu2022differential}, a method to improve ViT was applied to infant cry recognition by combining the original log-Mel spectrogram, first-order time-frame and frequency-bin differential log-Mels 3D features into ViT for infant cry recognition.
	
	\section{method}
	In this section, we describe the proposed model in detail. Our proposed model is shown in Fig. 1, which consists of three parts, namely CNN Block, T-Sa attention mechanism module, and LCT Block, these three modules will be introduced in detail next.
	\subsection{Overview of the model}
	To take full advantage of convolutional neural networks and transformer to model the speech sequences and use attention mechanism to enhance the features in time, space and channel, based on which our model is designed. As shown in Fig. 1, for a given input speech sequence, a series of preprocessing steps are performed. Specifically, we uniformly process different lengths of speech sequences into 1.8s, the longer sequences will be cropped into subsegments, and for shorter sequences, we process them using loop filling, after which MFCC features are extracted of speech as the input to the model. The local features of the speech first are obtained by a CNN block, where the irregular-sized time-frequency domain convolution is used to obtain the features in the time and frequency domains of the speech. The features are then enhanceed using a T-Sa attention mechanism block, which contains a bilstm attention module to model the features in the temporal order, followed by a spatial-channel attention mechanism to focus on the spatial and channel information. Finally, the global and local information of speech is learned interactively by an LCT Block, which enables the model to learn information at different scales. The three parts of the model in this paper are described in detail in the following sections.
	
	\subsection{CNN Block}
	For a large model such as transformer, if MFCC features are directly input into transformer, it will bring a large number of parameters. And since the dataset of speech emotion recognition is generally small, using transformer directly for feature learning will make the model difficult to converge. Therefore, we introduce a CNN Block to pre-learn the local features in speech. As shown in Figure 1, the CNN Block consists of a series of convolutional and pooling layers. For the input feature MFCC, the two dimensions of MFCC correspond to two dimensions in the temporal and frequency domains, respectively, for which we first use a pair of irregular convolutions to obtain the perceptual field in a specific range. For a convolution of size $3 \times 1$, we set the perceptual field in the time domain to 1, thus minimizing the effect in the time domain to learn information in the frequency domain, and for a convolution of size $1 \times 3$ which is a similar process, which in turn allows capturing a multi-scale representation in the temporal-frequency domain. The results are then fed to successive convolutional layers and maxpooling layers, which are used to further capture the local representation in speech, the batch normalization (BN), and relu activation function are applied after each convolutional layer.
	
	\subsection{T-Sa attention module}
	In this paper, inspired by \cite{zhang2021sa}, we adopt a novel attention module to combine temporal attention with spatial-channel attention, focusing on the temporal dynamics of speech features and spatial-channel information in the feature map, as shown in Fig. 2, which is divided into two parts before and after. In the former part, the attention model enhances the temporal information in the features by Bilstm to model the current features temporally, based on the fact that speech information is temporal information and the order of features in time is of importance. The latter part follows the spatial and channel attention, which is a common concern in existing attention, and efficiently combines the attention of the spatial and channel dimensions of speech feature maps through the Shuffle unit. T-Sa attention module enhances the features in the model through three dimensions and with a small number of parameters and achieves better results in SER. 
	
	\begin{figure*}[htbp]
		\centering
		\includegraphics[scale=0.055]{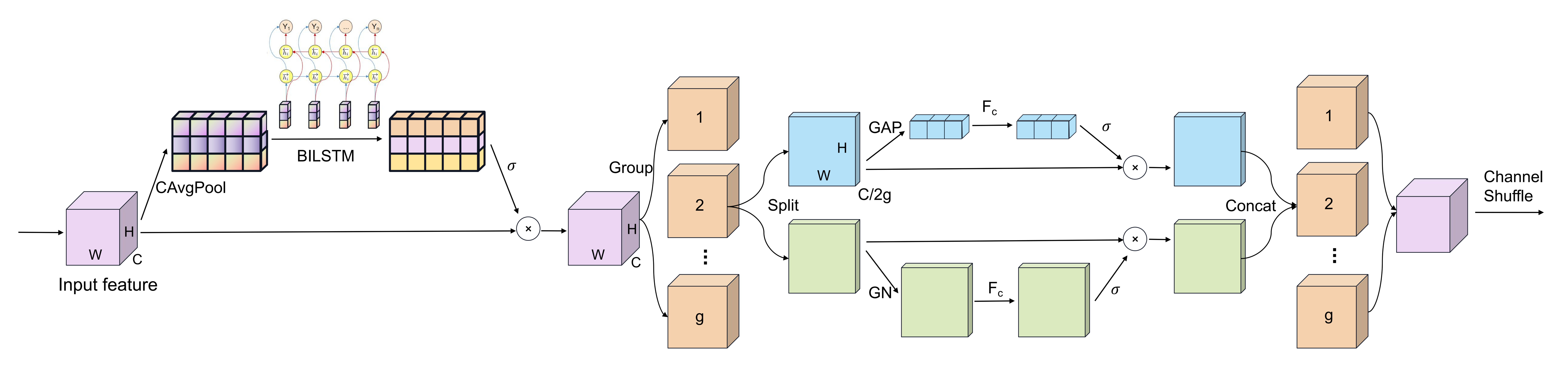}
		\caption{An illustration of our proposed T-Sa attention module consisting of two modules: Timing attention which enhances the current features temporally through BiLSTM, and Space-channel attention which enhances the features from the spatial and channel dimensions.}
	\end{figure*}
	
	\subsubsection{Timing attention}
	After Pre-processing and CNN Block for feature extraction of speech, giving the feature map size of $X \in R^{C\times H \times W}$ for the input T-Sa attention module, where $C$, $H$ and $W$ denote the number of channels, spatial height and width, respectively. To model the temporal attention in speech features, a recurrent neural network is used to model the speech information which is bilstm. The input of bilstm is a two-dimensional sequence while our speech feature map is three-dimensional, so what we need is to process the feature map $X$. If we directly reshape the feature map, the input bilstm will bring a great number of parameters number. Therefore, average pooling is used to reduce the dimensionality of the channels, and the specific calculation is:
	\begin{equation}
	X^{C Avg Pool}=\frac{1}{C} \sum_{i=1}^C X(i)
	\end{equation}
	
	After the feature passes through average pooling, the size of the feature map is $X^{C Avg Pool} \in R^{H \times W}$. Feature map is adjusted to $X^{C Avg Pool} \in R^{W \times H}$ by reshaping operation and then feeding it to the bilstm layer. By encoding long distances from front to back and from back to front, bilstm can better capture bidirectional feature dependencies. $X^{C Avg Pool}$ is encoded by bilstm as follows:
	\begin{equation}
		\overrightarrow{h}^{bilstm}=\overrightarrow{BILSTM}(X^{C Avg Pool})
	\end{equation}
	
	\begin{equation}
		\overleftarrow{h}^{bilstm}=\overleftarrow{BILSTM}(X^{C Avg Pool})
	\end{equation}
	Two LSTMs in bilstm process the sequence forward and backward, respectively, and then concate the outputs of the two LSTMs together:
	\begin{equation}
		H^{bilstm}=\operatorname{Concatenate }\left(\overrightarrow{h}^{bilstm} ; \overleftarrow{h}^{bilstm}\right)
	\end{equation}
	$H^{bilstm}$ is then applied sigmoid activation and multiplied with the input feature map using the residual scheme to output temporal attention:
	\begin{equation}
		X^{time}=\sigma\left(H^{bilstm}\right) \cdot X
	\end{equation}
	
	\begin{figure*}[htbp]
		\centering
		\includegraphics[scale=0.054]{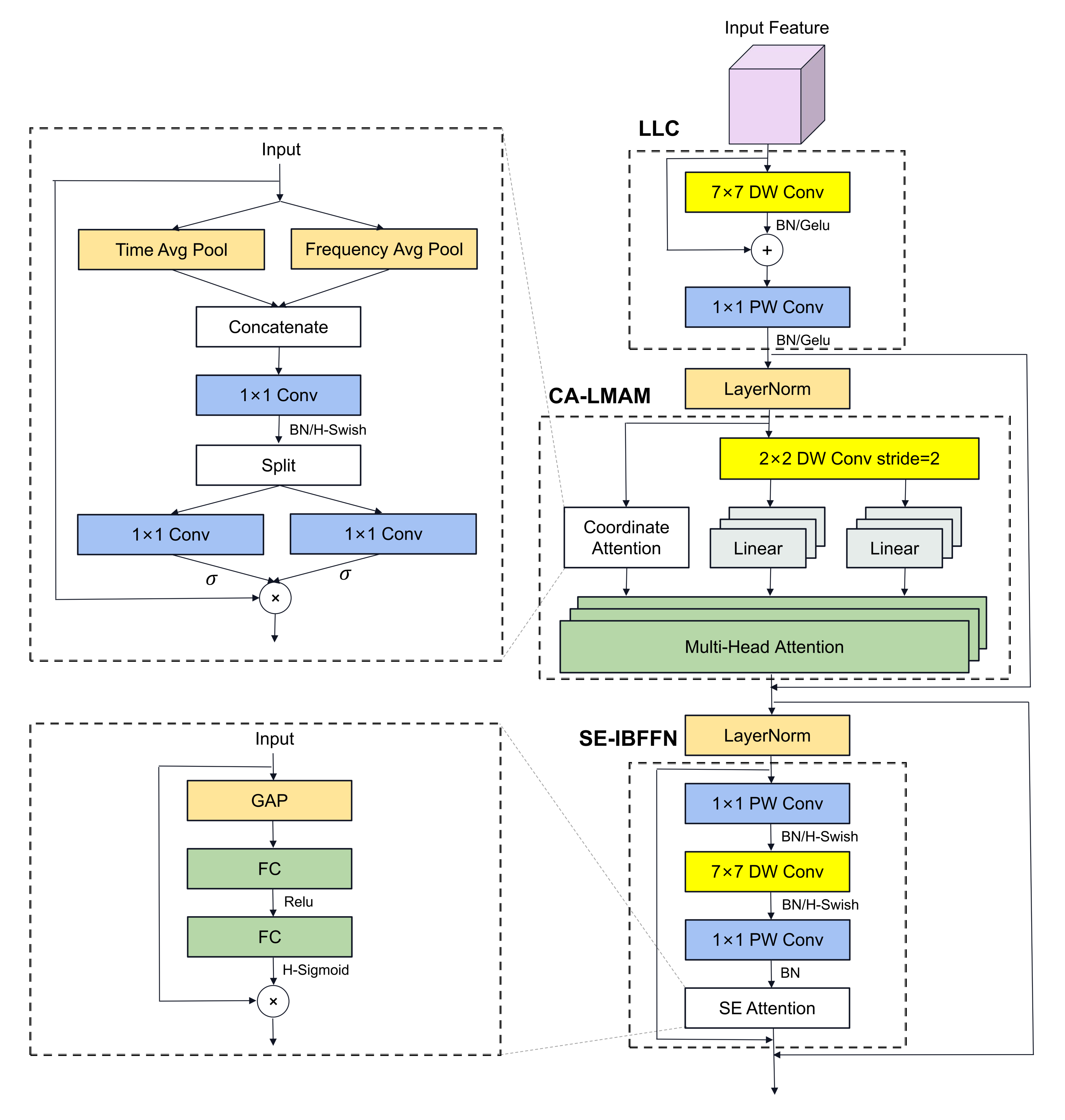}
		\caption{An illustration of our proposed LCT Block attention module consisting of three modules: i) LLC, which efficiently captures the local information of the features through a lightweight convolution module; ii) CA-LMAM, which captures the long-range dependencies in the features and enhance the time-frequency domain features of speech through a CA module; iii) SE-IBFFN, which introduces a nonlinear part through a feedforward network with inverse residuals to enhance the model the expressiveness of the model. }
	\end{figure*}
	
	\subsubsection{Space-channel attention}
	Spatial attention and channel attention are widely used in computer vision. Most methods transform and aggregate features in these two directions, such as SE \cite{hu2018senet}, CBAM \cite{woo2018cbam}, BAM \cite{park2018bam}, GCNet \cite{cao2019gcnet}, but these methods do not make full use of the correlation between space and channel which are not efficient. Therefore, we adopted SA-Net \cite{zhang2021sa} as our spatial-channel attention module.
	
	Given the output $X^{time}$ of the temporal attention module, the input is first divided into G groups, and the size of each sub-feature map is $X^{time^{\prime}} \in R^{C/G\times H \times W}$. Then, each group is split into two sub-branches in the channel dimension., $X^{spatial}$ and $X^{channel}$, one of which obtains spatial attention and the other obtains channel attention.
	
	For the channel attention branch, firstly, the global average pooling (GAP) operation is performed on the input of the branch to embed the global information. Then, a simple gating mechanism and sigmoid activation are used to perform adaptive learning of spatial features. A residual scheme is used to multiply the input channel branch feature map. The specific operation of the spatial attention module is as follows:
	\begin{equation}
		X^{channel^{\prime}}=\sigma\left(W_1 \cdot GAP(X^{channel})+b_1\right) \cdot X^{channel}
	\end{equation}
	$W_1$ and $b_1$ are learnable parameters.
	
	For the spatial attention branch, GN \cite{wu2018group} operation is first performed on the input of the branch to embed spatial statistical information, and then a simple gating mechanism and sigmoid activation are used to perform adaptive learning of features on the channel, and the residual scheme is used to multiply the input channel branch feature map. The specific operation of the channel attention module is as follows:
	\begin{equation}
		X^{spatial^{\prime}}=\sigma\left(W_2 \cdot GN(X^{spatial})+b_2\right) \cdot X^{spatial}
	\end{equation}
	Both $W_2$ and $b_2$ are learnable parameters, and then the outputs of the two branches are merged by concatenating them along the channel dimension:
	\begin{equation}
		X^{attention}=\operatorname{Concatenate}\left(X^{channel^{\prime}} ; X^{spatial^{\prime}}\right)
	\end{equation}
	
	The attention weights on space and channel are learned separately through two branches, and the corresponding residual scheme is multiplied with the respective input feature map to enhance the representations of space and channel. Finally, the channel shuffle \cite{ma2018shufflenet} is employed to facilitate communication of information between different groups along the channel dimension. The information of the features interacts in the channel dimension, and the size of the output feature map is the same as the initial input.
	
	\subsection{LCT Block}
	Inspired by \cite{guo2022cmt}, we proposed an LCT Block shown in Fig. 3, which consists of three modules: lightweight local convolution (LLC), coordinate attention-lightweight multi-head attention mechanism (CA-LMAM) and SE Inverted Bottleneck feed forward network (SE-IBFFN). Through the combination of convolution and transformer, LCT Block obtains the information of features from the local and the whole, which is more efficient and has fewer parameters than the traditional transformer. LLC is a lightweight convolution module, which efficiently obtains local information of features. The CA-LMAM module can capture the long-distance dependencies of features and enhance the information in the time-frequency domain through the Coordinate Attention (CA) \cite{hou2021coordinate} module. SE-IBFFN introduces a nonlinear part through a feedforward network with inverted residuals, which enhances the performance of the model and can further capture the local information of features. These three modules will be introduced in detail below
	\subsubsection{LLC}
	In order to make up for the lack of local information in trasnformer, a convolution module is used to obtain local information in speech which called LLC. Some work such as CMT \cite{guo2022cmt} also adopted a similar architecture, but the convolution module in CMT is relatively simple, only using a Depthwise convolution with residuals, which can not obtain effective local information. Inspired by \cite{trockman2022patches}, we used a large convolution kernel Depthwise convolution with residuals and a Pointwise convolution in the local feature extraction module, given an input feature $X \in R^{C\times H \times W}$ as follows:
	\begin{equation}
		LLC\left(X\right)=PWConv\left(DWConv\left(X\right)+X\right)
	\end{equation}
	The activation layer and batch normalization are omitted, and the same omission will be in subsequent formulas. $PWConv$ represents Pointwise convolution and $DWConv$ represents Depthwise. In $PWConv$ we used a large $7 \times 7$ convolution kernel is used to get a larger receptive field than a $3 \times 3$ convolution kernel. In addition, $PWConv$ has smaller parameters than $DWConv$ and ordinary convolution. Additionally, a residual structure is incorporated to address the issue of gradient dispersion.
	
	\subsubsection{CA-LMAM}
	In transformer, multi-head attention is usually used to make the model pay more attention to the more noteworthy part of itself, which can obtain long-distance dependence in speech features. Given the output $X \in R^{C\times H \times W}$ in the LLC module, the input of multi-head attention is query $Q$, key $K$, and value $V$, respectively. If the original features are directly input into multi-head attention, it will often bring a large amount of calculation. The amount of calculation is related to the size of the input features. Therefore, $2 \times 2$ $DWConv$ is used to downsample the parts of $K$ and $V$, as shown below:
	\begin{equation}
		K=DWConv\left(X\right)
	\end{equation}
	
	\begin{equation}
		V=DWConv\left(X\right)
	\end{equation}
	Where $K \in R^{C\times\frac{H} {2} \times \frac{W} {2}}$, $V \in R^{C\times\frac{H} {2} \times \frac{W} {2}}$, then the $H$ and $W$ dimensions are merged and input a linear layer respectively. Finally get $K^{\prime} \in R^{N^{\prime}\times C}$, $V^{\prime} \in R^{N^{\prime}\times C}$, where $N^{\prime}=\frac{H} {2} \times \frac{W} {2}$, $C$ is the dimension of the linear layer output.
	
	For $Q$, a CA module is used to enhance the time-frequency domain representation of speech features. This attention mechanism can obtain long-distance feature dependence along one direction and spatial dependence information in another direction. However, in speech, speech features have more special significance in the spatial dimension. The abscissa of speech features is the time axis, which represents the time domain information of speech while the ordinate of the speech feature represents the frequency information of the speech, so CA can better aggregate the dependent information in the time domain and frequency domain for speech features, which is more suitable for SER tasks.
	
	The specific operation of CA is shown in Fig. 3. Given the input $X \in R^{C\times H \times W}$, pooling is performed in the time domain and frequency domain respectively:
	\begin{equation}
		X^{T Avg Pool}=\frac{1}{W} \sum_{i=1}^W X(i)
	\end{equation}
	
	\begin{equation}
		X^{F Avg Pool}=\frac{1}{H} \sum_{j=1}^H X(j)
	\end{equation}
	$X^{T Avg Pool} \in R^{C\times H \times 1}$ and $X^{F Avg Pool} \in R^{C\times 1 \times W}$ are the feature maps after aggregation in two directions of time-frequency domain. Then, they concatenate and perform convolution operations to encode the information:
	\begin{equation}
		X^{tf}=Conv\left(\operatorname{concatenate}\left(X^{T Avg Pool} ; X^{F Avg Pool}\right)\right)
	\end{equation}
	Where $	X^{tf} \in R^{C/r\times 1 \times\left(W+H\right) }$, $r$ is the reduction rate of the channel, and then separated into two separate features $X^{t}$, $X^{f}$ along the spatial direction, where $X^{t} \in R^{C/r\times H \times 1}$, $X^{f} \in R^{C/r\times 1 \times W}$. The extraction of features is then performed by two separate convolutions, followed by activation using the $\sigma$ activation function, and then multiplied by the initial output to learn the more critical information in the feature:
	\begin{equation}
		{s}^t=\sigma\left(Conv\left(X^{t}\right)\right)
	\end{equation}
	
	\begin{equation}
		{s}^f=\sigma\left(Conv\left(X^{f}\right)\right)
	\end{equation}
	
	\begin{equation}
		Q=X\cdot{s}^t\cdot{s}^f
	\end{equation}
	We adjust the dimension of $Q$, eventually $Q^{\prime} \in R^{N\times C}$, where $N ={H\times W}$. 
	
	Eventually learning information about the model itself through multi-headed self-attention:
	\begin{equation}
		\operatorname{LMAM}(Q, K, V)=\operatorname{Softmax}\left(\frac{Q^{\prime} K^{\prime T}}{\sqrt{C}}+B\right) V^{\prime}
	\end{equation}
	Where $B$ is a learnable parameter, representing the relative position coding of multi-head self-attention, which is used to characterize the relative position relationship between tokens. It is more flexible than the traditional absolute position coding, making transformer better model the relative position information of speech features.
	
	\subsubsection{SE-IBFFN}
	In the original transformer, FFN is generally composed of two linear layers.  In this paper, inspired by \cite{howard2019searching}, a series of Pointwise convolution and Depthwise convolution make up our SE-IBFFN. Compared with the Transformer model traditionally composed of linear layers, this module can capture local information while learning channel information, and has a smaller number of convolution parameters than ordinary ones.
	
	The structure of SE-IBFFN is shown in Fig. 3. Given the input $X \in R^{C\times H \times W}$, the specific calculation process is as follows:
	\begin{equation}
		X^{conv}=PWConv\left(DWConv\left(PWConv\left(X\right)\right)\right)
	\end{equation}
	\begin{equation}
		SE\text{-}IBFFN\left(X\right)=SE\left(X^{conv}\right)+X
	\end{equation}
	Firstly, the Pointwise convolution operation is performed on the input, and the number of channels is expanded to 4 times of the original to increase the feature size of the channels. Then a $7 \times 7$ large convolution kernel $DWConv$ is used to obtain the local information in the feature, and the large convolution kernel can provide a larger receptive field without increasing too many parameters. The feature map is then restored to its original size using $PWConv$.
	
	Then, the SE module is to obtain the attention of the channel dimension of the feature map. Given the input $X \in R^{C\times H \times W}$ of SE, the specific calculation process is as follows:
	\begin{equation}
		SE\left(X\right)=\sigma\left(GAP\left(FC\left(FC\left(X\right)\right)\right)\right)\cdot X
	\end{equation}
	Compared with \cite{howard2019searching}, we put the SE module after $PWConv$, which makes the parameters of the model smaller. A residual structure is used to solve the problem of gradient dispersion.
	
	\section{experiments setup}
	In this section, we will introduce the dataset and experimental details used in our study, as well as the evaluation metrics used to assess the performance of our algorithm.
	\subsection{Corpus Description}
	To verify the performance of our proposed algorithm, performance tests on two benchmark databases are conducted, on which we will evaluate our algorithm.
	
	Actually, Interactive Emotional Dyadic Motion Capture (IEMOCAP) \cite{busso2008iemocap} is an action, multimodal, and multimodal database that contains data from 10 actors and actresses during an emotional binary interaction, with two speakers (one male and one female) speaking in each session. The IEMOCAP database has been annotated by several annotators with categorical labels and dimensional labels. The database combines discrete and dimensional sentiment models. In our work, the method used improvisational and scripted data, choose anger, happiness, neutral, sadness and excitement as basic emotions, and merge happy and excited into happy. We partitioned the IEMOCAP dataset into training and testing sets by randomly selecting 80\% and 20\% of the data, respectively.
	
	To valid our method robustness, we tested our method in other datasets. The Berlin Emotional Database (Emo-DB) \cite{burkhardt2005database} is a German emotional speech database recorded by the Technical University of Berlin. The database includes recordings of ten actors, comprising of five male and five female, who simulate seven emotions, including neutral, anger, fear, joy, sadness, disgust, and boredom, on ten sentences (five short and five long), resulting in a total of 535 speech recordings (233 male and 302 female). It has high emotional freedom, adopts 16 kHz sampling, and 16-bit quantization, and saves files in WAV format. It is a discrete emotional language database, and the excitation method is performance type.
	\subsection{Implementation Details}
	In the feature generation phase, the method uses the mel-frequency cepstrum coefficients (MFCCs) extracted by 26 Mel filters as the feature input. Meanwhile it divided each input speech into 1.8 seconds of speech segments, and the overlap between segments is 1.6 seconds, which can generate a large number of speech samples to solve the problem of scarcity of data set samples in SER. To obtain the prediction result for a sentence in the test set, we take the average of the prediction results of all speech segments within that sentence. Our model trained a total of 150 epochs, used the cross entropy criterion as the objective function, and used the Adam optimizer. The weight decay rate is $10^{-6}$, the learning rate and the mini-batch size are set to 0.001 and 128, respectively, and the multiplication factor 0.95 is exponentially decayed until the value reaches $10^{-6}$. Our experiment is carried out on Ubuntu 18.04 with a GeForce RTX 2080ti GPU, and we utilized Pytorch 1.7 as the training framework.
	
	In addition, we use the method of mixup \cite{zhang2017mixup} to train, so as to improve the generalization ability of the system. This method constructs new training samples and labels by linear interpolation, and effectively smoothes the discrete data space into continuous space. In our proposed model, we set $\alpha$ to 0.2 for best performance. 
	\subsection{Evaluation Metrics}
	In this section, we'll describe in detail the criteria we use to evaluate the performance of our algorithms. For various categories of performance in the dataset, Precision, Recall, and F1-score are the general metrics to measure their performance. First of all, four concepts will be introduced: True Positive (TP), False Positive (FP), True Negative (TN), and False Negative (FN), where TP means actual positive and predicted positive, FP means actual positive and predicted negative, TN means actual negative and predicted positive, and FN means actual negative and predicted negative. The Precision, Recall, and F1-score can be expressed as:
	\begin{equation}
		Precision=\frac{TP}{TP+FP}
	\end{equation}
	
	\begin{equation}
		Recall=\frac{TP}{TP+FN}
	\end{equation}
	
	\begin{equation}
		F1-score=\frac{precisio\times recall\times2}{precision+recall}
	\end{equation}
	
	To evaluate the overall performance of our model, weighted average accuracy (WA) and unweighted average accuracy (UA) will be used as evaluation metrics, where WA is the weighted average accuracy of different sentiment categories, and its weight is related to the number of sentiment categories, and UA is the average accuracy of different sentiment categories. The validity of the model is better evaluated in the context of unbalanced SER dataset samples. The calculation methods for these two metrics are as follows:
	\begin{equation}
		Acc_i=\frac{TP_i}{TP_i+FP_i}
	\end{equation}
	
	\begin{equation}
		WA=\frac{\sum_{i=1}^C N_i \times Acc_i}{\sum_{i=1}^C N_i}
	\end{equation}
	
	\begin{equation}
		UA=\frac{1}{C} \sum_{i=1}^C Acc_i
	\end{equation}
	Among them, $C$ represents the number of emotional categories, and $N_i$ represents the number of samples of class $i$.
	
	\section{results}
	In this section, we conduct extensive experiments to evaluate the performance of our method on two datasets. The section mainly compares our proposed model with state-of-the-art baselines, and then verify the effectiveness of our proposed modules through ablation experiments.

	\subsection{Comparison with State-of-the-Art}
	To compare with our proposed model, we evaluate the performance of the algorithm from the WA and UA perspective with a series of existing methods in IEMOCAP and Emo-DB datasets, as shown in Table \ref{tab:3} and Table \ref{tab:4}. The proposed model is compared with several commonly used speech emotion recognition models, including algorithms based on the combination of CNN and transformer \cite{hu2022multiple}, CNN-based algorithms \cite{liu2020speech}\cite{suganya2019speech}, LSTM-based algorithms \cite{latif2020deep}\cite{wang2020speech}, attention-based methods \cite{guo2021representation}\cite{li2021spatiotemporal}, and some other algorithms.
	
	\begin{table}[htbp]
	\centering
	\caption{COMPARISON ON IEMOCAP}
	\label{tab:3}  
	\renewcommand{\arraystretch}{1.3} 
	\begin{tabular}{ccc}
		\toprule 
		Comparative Methods & WA & UA \\
		\hline 
		Latif et al. \cite{9052467} & - & 68.8 \\
		Hu et al. \cite{hu2022multiple} & 69.73 & 70.11 \\
		Guo et al. \cite{guo2021representation} & 61.32 & 60.43 \\
		Gao et al. \cite{gao2021metric} & 70.34 & 70.82 \\
		Wang et al. \cite{wang2020speech} & 69.40 & 69.50 \\
		Latif et al. \cite{latif2020deep} & - & 64.10 \\
		Liu et al. \cite{liu2020speech} & 70.34 & 70.78 \\
		Dai et al. \cite{dai2019learning} & 65.40 & 66.90 \\
		\hline 
		\centering Proposed &\textbf{71.64} & \textbf{72.72} \\
		\bottomrule 
	\end{tabular}
\end{table}

\begin{table}[htbp]
	\centering
	\caption{COMPARISON ON EMO-DB}
	\label{tab:4}  
	\renewcommand{\arraystretch}{1.3} 
	\begin{tabular}{ccc}
		\toprule 
		Comparative Methods & WA & UA \\
		\hline 
		Tuncer et al. \cite{tuncer2021automated} & 90.09 & 89.47 \\
		Li et al. \cite{li2021spatiotemporal} & 83.30 & 82.10 \\
		Zhong et al. \cite{zhong2020exploration} & 85.76 & 86.12 \\
		Kerkeni et al. \cite{kerkeni2019automatic} & - & 86.22 \\
		Suganya et al. \cite{suganya2019speech} & - & 85.62 \\
		\hline 
		\centering Proposed & \textbf{90.65} & \textbf{89.51} \\
		\bottomrule 
	\end{tabular}
\end{table}
	
	To verify the performance of our proposed method, we compare it with other speech emotion recognition algorithms in IEMOCAP and Emo-DB datasets. The results are shown in Table \ref{tab:3} and Table \ref{tab:4}. 
	
	For IEMOCAP dataset, as shown in Table \ref{tab:3}, our proposed method outperforms other methods. In addition, compared with the simple splicing of transformer and CNN \cite{hu2022multiple}, our method achieves the best results through the ingenious design of local and global feature extraction. Compared to traditional spatial and channel attention \cite{guo2021representation}, temporal attention information makes it more competitive. In addition, for the CNN or LSTM-based method \cite{liu2020speech}\cite{wang2020speech}\cite{latif2020deep}\cite{dai2019learning}, our method introduces global information through a lightweight transformer module to bring more comprehensive features to the system, and also shows that our transformer module has a stronger ability to obtain global information than some capsule networks. Finally, our method is as competitive as the method using semi-supervised methods \cite{9052467} and pre-trained models \cite{gao2021metric}.
	
	\begin{figure*}[htbp]
		\centering
		\subfloat[]{\includegraphics[width=2.5in]{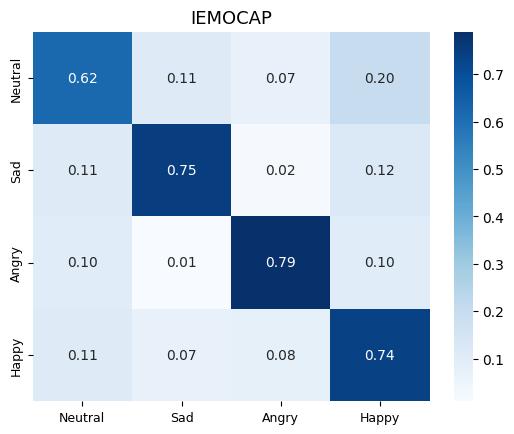}}%
		\hfil
		\subfloat[]{\includegraphics[width=2.72in]{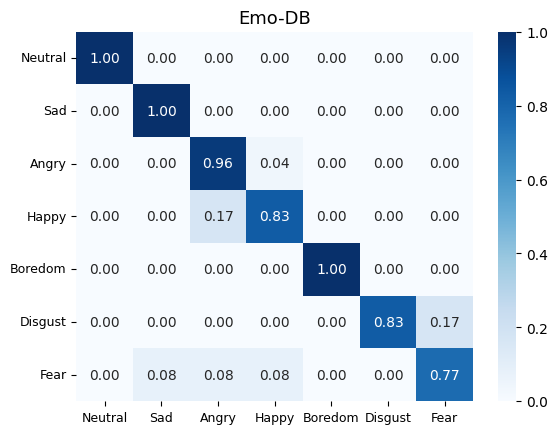}}%
		\caption{Visualization the confusion matrices of the proposed method: (a) Confusion matrix on IEMOCAP; (b) Confusion matrix on Emo-DB.}
	\end{figure*}
	
	For Emo-DB dataset, as shown in Table \ref{tab:4}, our proposed method outperforms several  methods. Similar to the performance in IEMOCAP dataset, our attention mechanism and local-global model have significant advantages over traditional attention and CNN-based models \cite{li2021spatiotemporal}\cite{zhong2020exploration}\cite{suganya2019speech}. In addition, Emo-DB dataset is a smaller dataset. Compared with non-deep learning feature extraction and feature learning methods \cite{tuncer2021automated}\cite{kerkeni2019automatic}, our overall end-to-end network based on deep learning also has a relatively better performance on this small dataset, indicating that our method still has excellent robustness on small datasets. 
	
	In summary, combining the performance of these two datasets proves the superiority of our proposed method.
	
	\subsection{Results and Analysis}
	In this section, Table \ref{tab:1} and Table \ref{tab:2} list the Precision, Recall, F1-score, and overall WA and UA for each sentiment category in IEMOCAP dataset and Emo-DB dataset, respectively. In addition, the confusion matrices are visualized of the two datasets in Fig. 4 and Fig. 5, where the diagonal indicates that the sentiment is correctly classified, and other locations indicate that the sentiment is misclassified as other sentiments. The darker the color in the grid while higher the accuracy. 
	
		\begin{table}[htbp]
		\centering
		\caption{Confusion matrix of the proposed model on IEMOCAP}
		\label{tab:1}  
		\renewcommand{\arraystretch}{1.3}
		\begin{tabular}{cccc}
			\toprule 
			Emotion & Precision & Recall & F1-score  \\
			\hline 
			Neural & 72.88 & 62.32 &67.19  \\
			Sad & 70.51 & 75.00 &72.68  \\
			Angry & 74.43 & 79.13 &76.71  \\
			Happy & 69.68 & 74.43 &71.98  \\
			\hline 
			WA &&71.64 & \\
			UA &&72.72 & \\
			\bottomrule 
		\end{tabular}
	\end{table}
	
	\begin{table}[htbp]
		\centering
		\caption{Confusion matrix of the proposed model on EMO-DB}
		\label{tab:2}  
		\renewcommand{\arraystretch}{1.3}
		\begin{tabular}{cccc}
			\toprule 
			Emotion & Precision & Recall & F1-score  \\
			\hline 
			Neural & 86.67 & 100.00 &92.86  \\
			Sad & 94.44 & 100.00 &97.14  \\
			Angry & 88.00 & 95.65 &91.67  \\
			Happy & 83.33 & 83.33 &83.33  \\
			Boredom & 100.00 & 93.75 &96.77  \\
			Disgust & 100.00 & 76.92 &86.96  \\
			Fear & 83.33 & 76.92 &80.00  \\
			\hline 
			WA &&90.65 & \\
			UA &&89.51 & \\
			\bottomrule 
		\end{tabular}
	\end{table}
	
	For IEMOCAP dataset, as shown in Table \ref{tab:2} and Fig. 4, our proposed model achieves good results on this dataset and has good accuracy for all four emotions, especially sadness, anger and happiness. Among these four emotions, anger has the highest recognition accuracy, while neutral has the lowest recognition accuracy. Sadness, anger and happiness will be misjudged as neutral in some cases. The result is similar to that in \cite{hou2021multi}, which also verifies that neutral emotion is a defect of expression. This emotion is easily expressed as other emotions, which makes the model misjudged. Therefore, neutral emotions are easily confused with other emotions. 
	
	For Emo-DB dataset, as shown in Table \ref{tab:2} and Fig. 5, our proposed model also achieved good results on this dataset, and has good accuracy for seven emotions. The three emotions achieved 100\% accuracy, and Angry also achieved 96\% accuracy. In addition, the wrong judgment in happiness is anger. We believe that this is because these two emotions have similar arousal, while the wrong judgment in disgust is fear. Finally, the accuracy of fear is lower than that of other emotion categories, but it also achieves relatively good results. In some cases, fear is easily misjudged as sad, angry and happy, this is because these four emotions have a high degree of arousal.
	
	\begin{table}[htbp]
		\centering
		\caption{Performance with different experimental settings}
		\label{tab:5} 
		\renewcommand{\arraystretch}{1.3} 
		\begin{tabular}{ccc}
			\toprule 
			Models & WA & UA \\
			\hline 
			W/o T-Sa & 69.92 & 71.18 \\
			W/o lstm attention & 67.84 & 69.62 \\
			W/o LCT & 67.12 & 68.93 \\
			W/ Conv-LCT & 68.29 & 69.98 \\
			W/o CA & 69.29 & 70.12 \\
			W/o SE & 69.92 & 71.31 \\
			\hline 
			Proposed & \textbf{71.64} & \textbf{72.72} \\
			\bottomrule 
		\end{tabular}
	\end{table}
	\subsection{Ablation Study}
	To explore the role of each part of our proposed model, Table \ref{tab:5} shows the results of a series of ablation experiments. The following are the modules for comparison.
	\begin{itemize}
		\item W/o T-Sa: This module removes the T-Sa module from our model, using only the CNN Block and LCT Block sections. 
		\item W/o lstm attention: This module removes the temporal attention lstm attention part in our T-Sa module, and the other parts are retained.  
		\item W/o LCT: This module removes the LCT module in our model and only uses the CNN Block and T-Sa parts.  
		\item W/ Conv-LCT: This module replaces the LLC module in our LCT module with a $3 \times 3$ convolution, and the rest is preserved.
		\item W/o CA: This module removes the CA portion of our LCT module and preserves the rest. 
		\item W/o SE: This module removes the SE Attention part of our LCT module, and the other parts are retained.
	\end{itemize}
	
	It can be seen from the table that when using the T-Sa module, our method has an absolute improvement of 1.2\% and 1.54\% in WA and UA, indicating that our attention mechanism module has a very significant effect and can aggregate the noteworthy parts of the features. In addition, when the temporal attention is removed, the model effect is also greatly reduced, indicating that temporal attention plays a non-important role in our model to enhance the temporal information in speech. We also directly removed the LCT module, which caused the performance of the model to be reduced by 4.52\% and 3.79\% on WA and UA, it clearly shows the importance of introducing global information into our LCT module. 
	
	For our LCT part, our experiments also verified the role of different modules within the LCT. Firstly, the LLC part is replaced with a $3 \times 3$ ordinary convolution, which reduces the performance of the model by 3.35\% and 2.74\% on WA and UA, and the number of parameters has also been improved. This shows that the wider receptive field brought by our large convolution kernel LLC module is very important, and it does not bring a larger number of parameters. In order to verify the role of our CA module, we removed the CA module, which caused the performance of the model to decrease by 2.35\% and 2.6\% on WA and UA, indicating that the time-frequency domain representation of the speech features enhanced by the CA module is of vital importance. Finally, the SE Attention part in our SE-IBFFN is removed, which reduces the performance of the model by 1.72\% and 1.41\% on WA and UA. It also proves that using the SE module to obtain the attention of the feature map channel dimension can improve the performance of the model to recognize emotions.
	
	\begin{table}[htbp]
		\centering
		\caption{Comparison of Model Parameters and Accuracy}
		\label{tab:6} 
		\renewcommand{\arraystretch}{1.3} 
		\begin{tabular}{cccc}
			\toprule 
			Models &Params& WA & UA \\
			\hline 
			W/ Conv-LCT &1,404,922 & 68.29 & 69.98 \\
			W/ConvFFN-LCT-L &10,390,522 & 70.46 & 72.14 \\
			W/ConvFFN-LCT-S &2,526,202 & 68.83 & 69.71 \\
			\hline 
			W/Transformer &1,357,810 & 69.38 & 70.43 \\
			W/MobileVitv1 block-depth2\cite{mehta2021mobilevit} &2,036,458 & 70.01 & 70.98 \\
			W/MobileVitv1 block-depth3\cite{mehta2021mobilevit} &2,726,506 & 66.58 & 67.30 \\
			W/MobileVitv2 block-depth2\cite{mehta2022separable} &937,898 & 68.29 & 69.68 \\
			W/MobileVitv2 block-depth3\cite{mehta2022separable} &1,086,634 & 67.12 & 69.79 \\
			W/MobileVitv3 block-depth2\cite{wadekar2022mobilevitv3} &\textbf{661,162} & 69.20 & 70.14 \\
			W/MobileVitv3 block-depth3\cite{wadekar2022mobilevitv3} &884,010 & 68.47 & 69.73 \\
			\hline 
			Proposed  &1,031,674 & \textbf{71.64} & \textbf{72.72} \\
			\bottomrule 
		\end{tabular}
	\end{table}
	
	\subsection{Model Efficiency Analysis}
	In order to explore the size and efficiency of the proposed model, Table \ref{tab:6} presents a comparison of the parameter counts and accuracy between our model and other models. The following are the modules we designed for comparison. 
	\begin{itemize}
		\item W/ Conv-LCT: This module replaces the LLC module in our LCT module with a $3 \times 3$ convolution and the other parts are retained. 
		\item w/ ConvFFN-LCT-L: This module replaces the first PW convolution and DW convolution in the SE-IBFFN module of our LCT module with a $7 \times 7$ convolution and the other parts are retained.
		\item W/ ConvFFN-LCT-S: This module replaces the first PW convolution and DW convolution in the SE-IBFFN module of our LCT module with a $3 \times 3$ convolution and the other parts are retained.  
		\item W/Transformer: This module replaces CA-LMAM and SE-IBFFN with traditional transformer, which is the same as the setting of LCT. 
		\item W/ MobileVit block: This module replaces our LCT module with the MobileVit block of various series of Mobile Vit, with an input channel size of 128 and a depth of 2 or 3. The intermediate layer size of the MLP is set to 4 times of the input channel size for all versions. For v1 and v2, the number of intermediate layers is 192, and for v3, the number of intermediate layers is 128.
	\end{itemize}
	
	From the  Table \ref{tab:6}, it can be seen that when the LLC module in LCT is replaced with a regular convolution, the number of parameters increases and the accuracy is lower than that of the LLC module. This indicates that LLC has better accuracy with fewer parameters. Additionally, when the first PW and DW convolutions in SE-IBFFN are replaced with regular convolutions, the number of parameters increases significantly, and the accuracy is still lower than the original model. We also reduced the size of the convolution kernel to  $3 \times 3$, which improved the number of parameters, but it is still larger than the original SE-IBFFN, and the accuracy decreased. This suggests that the SE-IBFFN module did not bring a huge number of parameters while using a larger convolution kernel to achieve a larger receptive field, and the accuracy is still excellent.
	
	Additionally, we attempted to replace CA-LMAM and SE-IBFFN in LCT with traditional transformers, and experimental results showed that the traditional transformer has a larger number of parameters and a decrease in accuracy compared to LCT, with a decrease of 2.23\% and 2.29\%, respectively. The entire LCT module was also replaced with other transformer models, including the MobileVit series which combines CNN and Vit and is a lightweight transformer model. The MobileVit block in each series was used to replace LCT, and the results showed that the parameter size of the MobileVitv3 block is smaller than LCT, but there is still a gap in accuracy. The performance of the two-layer MobileVitv1 block was the best, but it still did not reach the accuracy of LCT. Additionally, we found that models with a depth of 2 performed better than those with a depth of 3, due to the difficulty of fitting to small SER datasets as the number of parameters increases.
	
	\section{Conclusion}
	In this study, to better model local and global features of speech signals at different levels of granularity in SER and capture temporal, spatial and channel dependencies in speech signals, we propose a Speech Emotion Recognition network based on CNN-Transformer and multi-dimensional attention mechanisms. The network consists of three modules. First, a CNN block is used to model time-frequency domain information in speech, capturing preliminary local information in speech. Second, we propose a T-Sa network to model the emotional expression context of features over time and efficiently fuse the spatial and channel dimensions of speech feature maps through Shuffle units. Finally, to efficiently fuse local information and long-distance dependencies in speech, we propose an LCT module that uses lightweight convolutional modules and introduces Coordinate Attention into multi-head self-attention. This allows for the fusion of features at different levels of granularity while enhancing information in the time-frequency domain of features without introducing a high number of parameters.
	
	In future work, in addition to MFCC features, we will try more hierarchical speech features and combine current CNN and transformer structures to improve the performance of Speech Emotion Recognition from multiple feature dimensions.
	
	\section*{Acknowledgments}
	This work was supported by the National Natural Science Foundation of China (Grant No. 62001173), the Project of Special Funds for
	the Cultivation of Guangdong College Students’ Scientific and Technological Innovation ("Climbing Program" Special Funds) (Grant No.
	pdjh2022a0131, pdjh2023b0141).
	
	\bibliographystyle{IEEEtran}  
	\bibliography{IEEEabrv,ref}

    \begin{IEEEbiography}[{\includegraphics[width=1in,height=1.25in,clip,keepaspectratio]{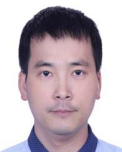}}]{Xiaoyu Tang}
		(Member, IEEE) received the B.S. degree from South China Normal University in 2003 and the M.S. degree from Sun Yat-sen University in 2011. He is currently pursuing the Ph.D. degree with South China Normal University. He is working with the School of Physics and Telecommunication Engineering, South China Normal University, where he engaged in information system development. His research interests include machine vision, intelligent control, and the Internet of Things. He is a member of the IEEE ICICSP Technical Committee.
	\end{IEEEbiography}
	
    \begin{IEEEbiography}[{\includegraphics[width=1in,height=1.25in,clip,keepaspectratio]{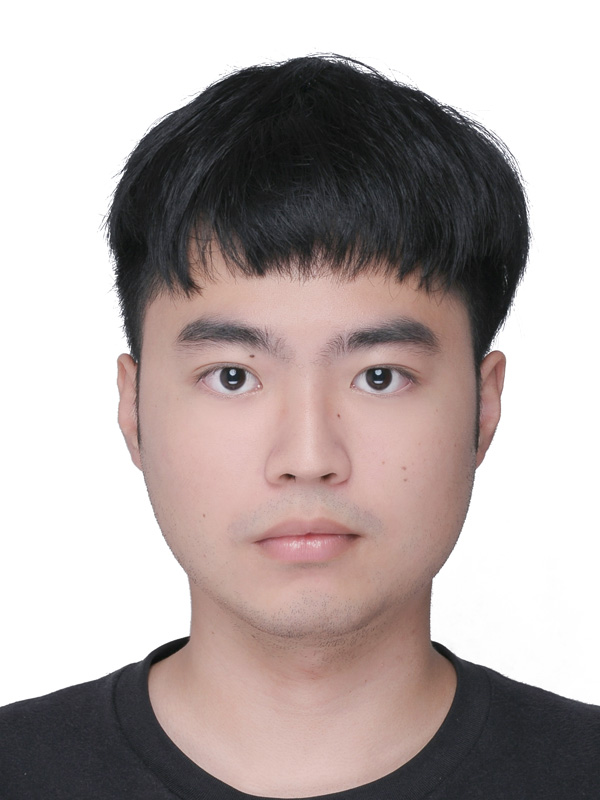}}]{Yixin Lin}
		received the B.Eng. degree from the School of Physics and Telecommunication Engineering, South China Normal University, in 2021, where he is currently pursuing the M.E. degree with the Department of Electronics and Information Engineering. His research interests include artificial intelligence and speech emotion recognition.
	\end{IEEEbiography}

    \begin{IEEEbiography}
    [{\includegraphics[width=1in,height=1.25in,clip,keepaspectratio]{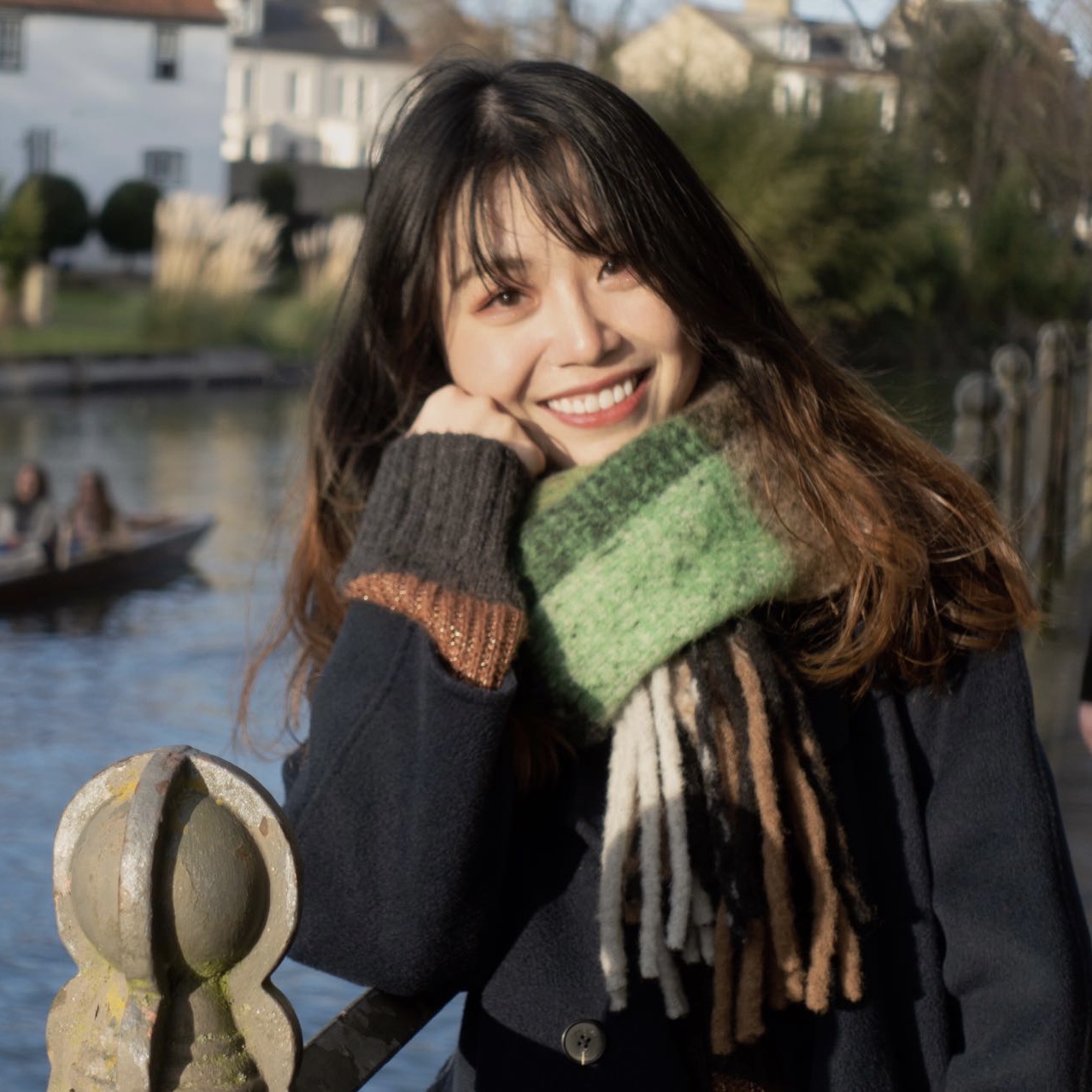}}]{Ting Dang}
		is a Senior Research Scientist at Nokia Bell Labs, Cambridge, UK. Before joining Nokia, she worked as a Senior Research Associate at the University of Cambridge. Dr. Dang earned her Ph.D. degree from the University of New South Wales in Sydney, Australia, and holds MEng and BEng degrees in Signal Processing from the Northwestern Polytechnical University in China. Her primary research interests are on exploring the potential of audio signals (e.g., speech) for mobile health, i.e., automatic disease and mental state prediction and monitoring (e.g., COVID-19, emotion) via mobile and wearable audio sensing. Her work aims to develop generalised and interpretable machine learning models to improve healthcare delivery. She served as the (senior) program committee and invited reviewer for more than 30 top-tier conferences and journals, such as AAAI, NeurIPS, ICASSP, IEEE TAC, IEEE TASL etc. She has won the Asian Dean’s Forum (ADF) Rising Star Women in Engineering Award 2022, IEEE Early Career Writing Retreat Grant 2019 and ISCA Grant 2017.
	\end{IEEEbiography}

	\begin{IEEEbiography}
		[{\includegraphics[width=1in,height=1.25in,clip,keepaspectratio]{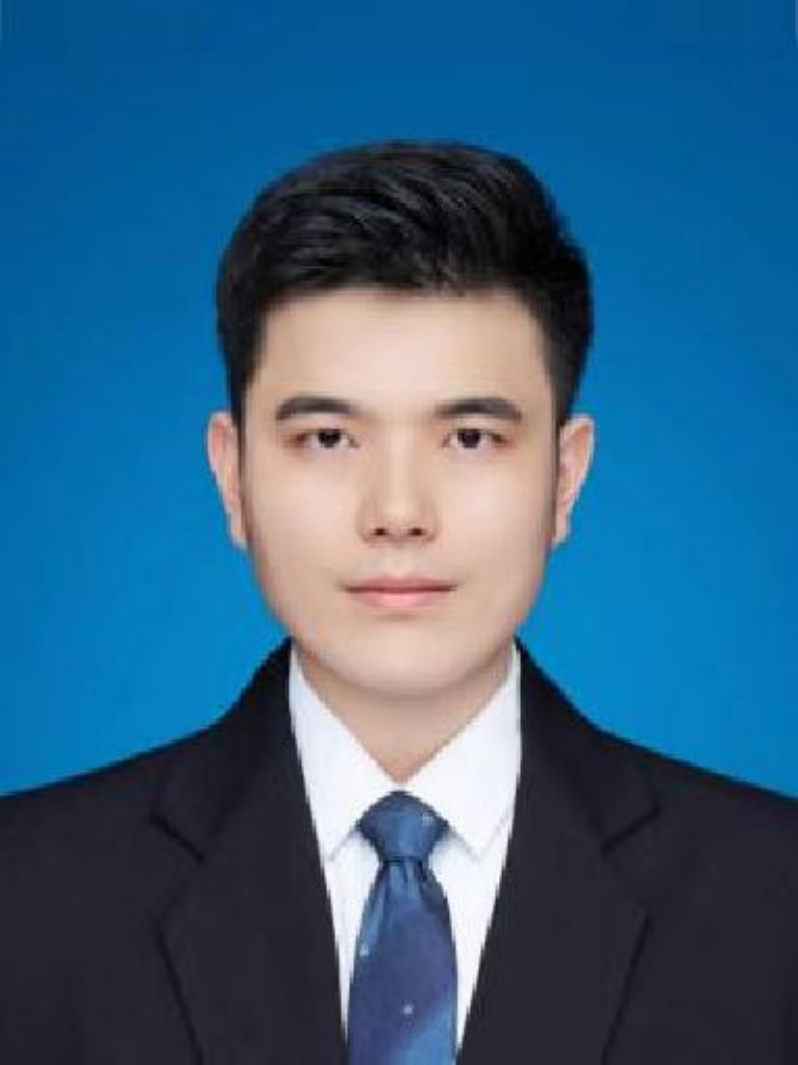}}]{Yuanfang Zhang}
		is currently principal researcher at Autocity (Shenzhen) autonomous driving Co.,ltd. He got his dual Ph.D. degrees with the School of Computer Science at Northwestern Polytechnical University, Shaanxi Province, China and Faculty of Engineering and IT, University of Technology Sydney, Australia. His current research focuses on unmanned vehicle sensing technology, multimodal learning methodology, comprehensive computer vision.
	\end{IEEEbiography}

	\begin{IEEEbiography}
		[{\includegraphics[width=1in,height=1.25in,clip,keepaspectratio]{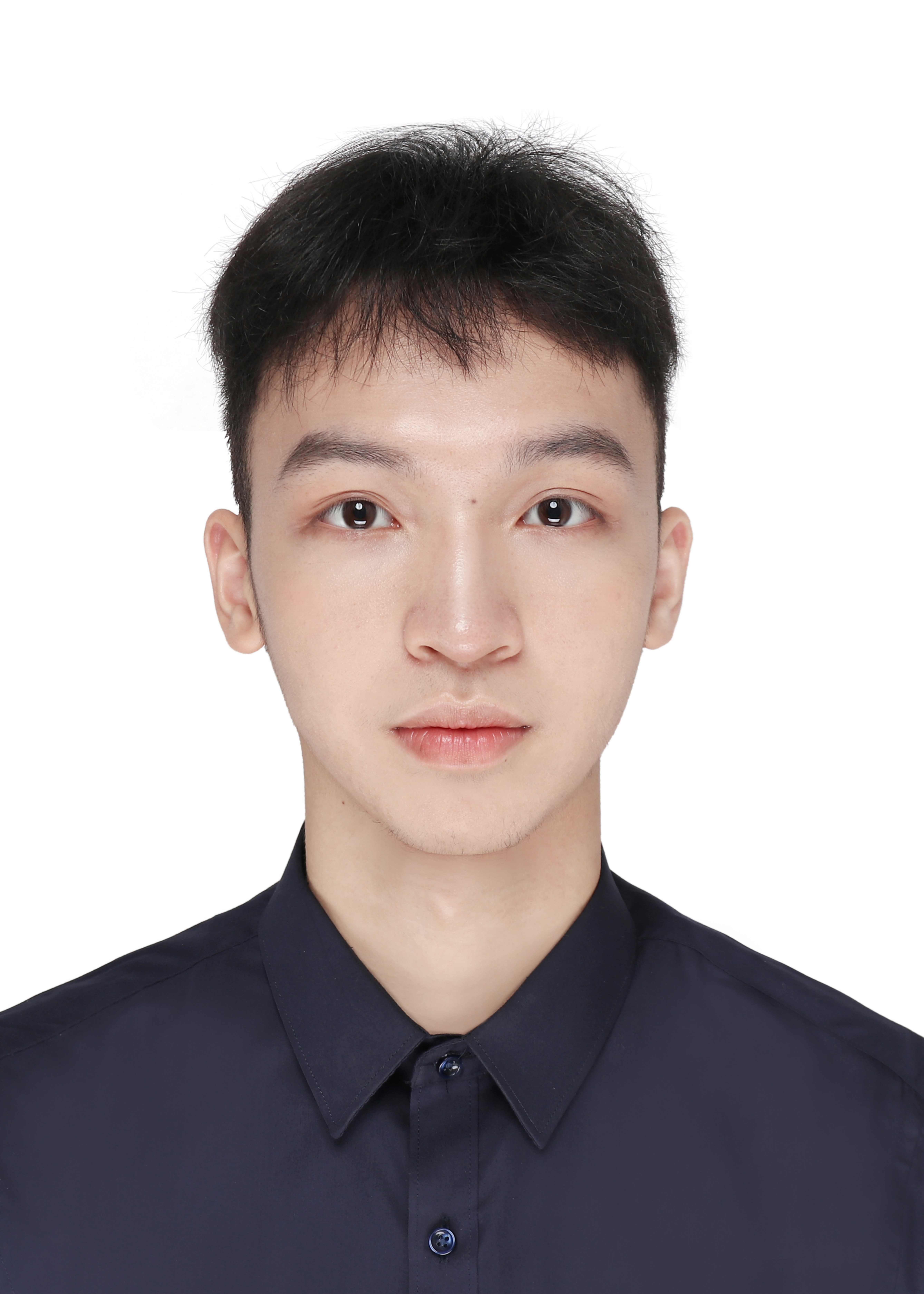}}]{Jintao Cheng}
		received his bachelor's degree from the school of Physics and Telecommunications Engineering, South China Normal University in 2021. His research is computer vision, SLAM and deep learning.
	\end{IEEEbiography}
	
\end{document}